\documentclass[amsmath,floatfix,twocolumn,superscriptaddress]{revtex4-1}
\usepackage{subfigure}
\usepackage{siunitx}
\usepackage{amssymb}
\usepackage{amsmath}
\usepackage{graphicx}
\usepackage{array}
\usepackage{dcolumn}
\usepackage{psfrag}
\usepackage{bm}
\usepackage{color}

\begin{document}

\title{Chalcogen Vacancies Rule Charge Recombination in Pnictogen Chalcohalide Solar-Cell Absorbers}

\author{Cibrán López}
    \affiliation{Departament de Física, Universitat Politècnica de Catalunya, 08034 Barcelona, Spain}
    \affiliation{Barcelona Research Center in Multiscale Science and Engineering, Universitat Politècnica de Catalunya, 08019 Barcelona, Spain}

\author{Seán R. Kavanagh}
    \affiliation{Harvard University Center for the Environment, Cambridge, Massachusetts 02138, United States}

\author{Pol Benítez}
    \affiliation{Departament de Física, Universitat Politècnica de Catalunya, 08034 Barcelona, Spain}
    \affiliation{Barcelona Research Center in Multiscale Science and Engineering, Universitat Politècnica de Catalunya, 08019 Barcelona, Spain}

\author{Edgardo Saucedo}
    \affiliation{Barcelona Research Center in Multiscale Science and Engineering, Universitat Politècnica de Catalunya, 08019 Barcelona, Spain}
    \affiliation{Department of Electronic Engineering, Universitat Politècnica de Catalunya, 08034 Barcelona, Spain}

\author{Aron Walsh}
    \affiliation{Thomas Young Centre and Department of Materials, Imperial College London, Exhibition Road, London SW7 2AZ, UK}
    \affiliation{Department of Physics, Ewha Womans University, 52 Ewhayeodae-gil, Seodaemun-gu, Seoul 03760, South Korea}

\author{David O. Scanlon}
    \affiliation{School of Chemistry, University of Birmingham, Birmingham B15 2TT, UK}

\author{Claudio Cazorla}
    \affiliation{Departament de Física, Universitat Politècnica de Catalunya, 08034 Barcelona, Spain}
    \affiliation{Barcelona Research Center in Multiscale Science and Engineering, Universitat Politècnica de Catalunya, 08019 Barcelona, Spain}

\maketitle

{\bf
Pnictogen chalcohalides (MChX, M = Bi, Sb; Ch = S, Se; X = I, Br) represent an emerging class of nontoxic photovoltaic absorbers, valued for their favorable synthesis conditions and excellent optoelectronic properties. Despite their proposed defect tolerance, stemming from the antibonding nature of their valence and conduction bands, their experimentally reported power conversion efficiencies remain below 10\%, far from the ideal Shockley-Queisser limit of 30\%. Using advanced first-principles calculations and defect sampling techniques, we uncover a complex point-defect landscape in MChX materials, exemplified by BiSeI. Previously overlooked selenium vacancies are identified as critical nonradiative charge-recombination centers, which exist in high concentrations and, although exhibit modest capture coefficients, can reduce the maximum power conversion efficiency of BiSeI down to 24\%. We argue that such detrimental effects can be mitigated by cation-poor synthesis conditions and strategic anion substitutions. Building on these insights, and supported by further simulations, we predict BiSBr to be a more defect-tolerant light absorber. This study not only identifies efficiency-limiting factors in MChX  but also provides a roadmap for their improvement, paving the way for next-generation solution-processed chalcogenide photovoltaics.
}
\\

Pnictogen chalcohalides (MChX) have garnered significant attention as promising solar-cell absorber materials due to their nontoxicity, low synthesis temperatures (below $300^\circ$C) \cite{Cano2023,Li2024}, optimal bandgaps ranging from $1.0$ to $2.0$~eV \cite{Ghorpade2023,He2023}, and exceptional thermodynamic stability \cite{Nie2019,LiY2023}. Their electron affinities and ionization potentials also align well with established charge transport layers \cite{Tiwari2019,Guo2023}. Additionally, their wide bandgap range and high optical absorption coefficients extend their applicability to multi-junction solar cell devices, which can potentially exceed the power-conversion efficiencies of conventional single-junction solar cells. These advantageous properties underscore the potential of MChX for next-generation solution-processed solar energy technologies.

Notably, a gap persists between theoretical predictions, which characterize MChX as exceptional photoabsorbers \cite{Lopez2024,Ganose2016}, and experimental findings, which report suboptimal MChX photovoltaic performance \cite{Cano2023,Nie2021,Tiwari2019}. With power conversion efficiencies (PCE) currently below $10\%$, far from the ideal detailed balance limit of $30\%$ \cite{Shockley1961}, MChX face significant barriers to commercial viability. This poor PCE is likely due to reduced carrier lifetimes and nonradiative electron-hole recombination resulting from deep recombination-active defect levels \cite{Kavanagh2024-2}.

MChX semiconductors exhibit antibonding states at the valence band maximum (VBM) and conduction band minimum (CBM), along with high dielectric constants and high charge-carrier mobilities \cite{Ran2018}, similar to lead-halide perovskites \cite{Walsh2017,Huang2021}. These features are believed to promote the formation of shallow defect energy levels near the band edges, rather than deep defect levels \cite{Zakutayev2014}, indicating potential defect tolerance \cite{Ran2018,Zhang1998}. However, the contrast between this suggested defect tolerance and the observed photovoltaic underperformance suggests the need for a comprehensive investigation of defect chemistry in MChX. This analysis is critical to identify the most detrimental defects and develop effective defect-passivation strategies \cite{Savory2019}.

In this study, we present a theoretical investigation of the defect chemistry in MChX, focusing on the representative compound BiSeI. Employing advanced first-principles calculations and defect sampling techniques, we perform a comprehensive analysis of intrinsic point defects, including vacancies, antisites, and interstitials. Among these, Se vacancies (V\textsubscript{Se}) are identified as the most detrimental defects, significantly undermining photovoltaic efficiency by facilitating nonradiative trap-mediated charge recombination.  
\\

\begin{figure*}[t]
    \centering
        \includegraphics[width=\textwidth]{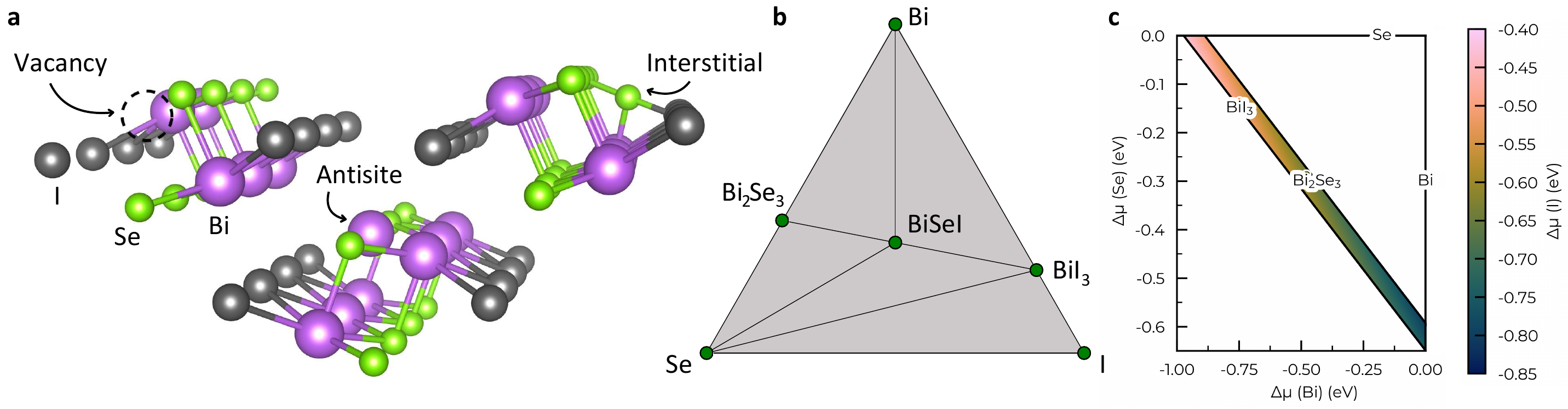}
        \caption{\textbf{Structural and phase stability properties of BiSeI.}
        \textbf{a.}~BiSeI crystal structure (orthorhombic, $Pnma$) characterized by columnar motifs held together by weak van der Waals forces. Point defects considered in this study: vacancies, antisites, and interstitials. Bi, Se and I atoms are represented with purple, green and grey spheres, respectively. 
        \textbf{b.}~Convex-hull surface of BiSeI calculated with DFT methods. BiSeI is predicted to be thermodynamically stable against separation into secondary phases because its formation enthalpy is negative relative to the convex-hull surface.
        \textbf{c.}~Chemical stability region, delimited by Se-poor \((\mu_{\rm Bi}, \mu_{\rm Se}, \mu_{\rm I}) = (0, -0.65, -0.75)\)~eV and Bi-poor conditions \((\mu_{\rm Bi}, \mu_{\rm Se}, \mu_{\rm I}) = (-0.97, 0, -0.42)\)~eV (Supplementary Table~1).}
        \label{Figure: BiSeI structure}
\end{figure*}

\textbf{Computational framework.}~Semiconducting BiSeI crystallizes in an orthorhombic phase (space group $Pnma$) characterized by one-dimensional columns held together by weak van der Waals forces (Fig.~\ref{Figure: BiSeI structure}a), closely resembling the structure of pnictogen chalcogenides (e.g. Sb\textsubscript{2}Se\textsubscript{3} or Bi\textsubscript{2}Se\textsubscript{3} \cite{Zhou2015}). Our density functional theory (DFT) geometry optimizations yield lattice parameters that are in very good agreement with the available experimental data \cite{Lopez2024} (i.e., $a^{\rm DFT} = 4.27$~Å, $b^{\rm DFT} = 9.04$~Å and $c^{\rm DFT} = 11.28$~Å, to be compared with $a^{\rm expt} = 4.22$~Å, $b^{\rm expt} = 8.70$~Å and $c^{\rm expt} = 10.58$~Å). According to our DFT calculations, BiSeI is thermodynamically stable against phase separation into Bi\textsubscript{2}Se\textsubscript{3} and BiI\textsubscript{3} (i.e., $\Delta H = -0.01$~eV/atom referred to the convex hull surface, Fig.~\ref{Figure: BiSeI structure}b), with the range of stable chemical potentials for this system given in Fig.~\ref{Figure: BiSeI structure}c. These findings are consistent with experimental observations \cite{Cano2023,Li2024}, but differ from Materials Project predictions \cite{Jain2013} that neglect long-range dispersion forces (Methods).

Crystalline defects can be broadly classified into point and extended defects. Point defects (Fig.~\ref{Figure: BiSeI structure}a) include vacancies, where an atom is removed from the lattice (e.g., V\textsubscript{Se}), antisites, where an atom is replaced by another of a different species (e.g., Bi\textsubscript{Se}), and interstitials, where an atom occupies a nonequilibrium lattice site (e.g., Se\textsubscript{i}). Higher-dimensional defects, such as grain boundaries, dislocations, and precipitates, may also form in materials \cite{Ball2016}. However, recent experimental studies indicate that these defects are not prevalent in MChX \cite{Cano2023}. Furthermore, chain-like structures are likely to produce grain boundaries that are charge-recombination inactive \cite{McKenna2021,Williams2020,Zhou2015}.  Consequently, this computational work focuses on point defects. 

Computational approaches for studying point defects in crystals are well-established, relying on accurate first-principles energy calculations combined with exhaustive exploration of the defect local environment \cite{Mosquera-Lois2022,Freysoldt2009,Kumagai2014}. For this work, we employed the supercell approach, which involves modeling point defects within sufficiently large supercells to minimize spurious interactions (Methods). We systematically analyzed all possible vacancy and antisite defects, considering both neutral and charged states. For interstitial defects, we initially evaluated their neutral states by sampling all possible sites obtained from Voronoi analysis (Supplementary Fig.~1). Interstitials with sufficiently low formation energies in the neutral state were further analyzed considering multiple charged states.
The ShakeNBreak \cite{Mosquera-Lois2022,Mosquera-Lois2021} defect structure-search approach was employed, revealing numerous significant energy-lowering reconstructions --consistent with observations in similar low-dimensional chalcogenide systems \cite{Wang2023,Wang2025,Mosquera-Lois2024}. 

First-principles calculations based on density functional theory (DFT) \cite{Cazorla2017} were performed using the VASP code \cite{Kresse1996}. To address the limitations of semilocal functionals \cite{Freysoldt2014}, we employed the range-separated hybrid functional HSEsol, which is based on the Perdew–Burke–Ernzerhof exchange-correlation functional revised for solids \cite{Perdew2008,Heyd2003,Schimka2011}. Long-range dispersion interactions were taken into account through the van der Waals D3 correction scheme \cite{Grimme2010}. Additionally, spin-orbit coupling (SOC) effects, which are particularly relevant for Bi-based compounds, were explicitly included \cite{Heyd2003,Krukau2006}. All defective atomic structures were fully optimized at the HSEsol+D3+SOC level, a methodology shown to accurately reproduce experimental results for materials similar to MChX \cite{Lopez2024,Pan2018}. The \verb!doped! simulation package \cite{Kavanagh2024} was used to generate defect structures and calculation inputs, determine chemical potential limits, and analyze the defects simulation results.
\\

\begin{figure*}[t]
    \centering
        \includegraphics[width=\textwidth]{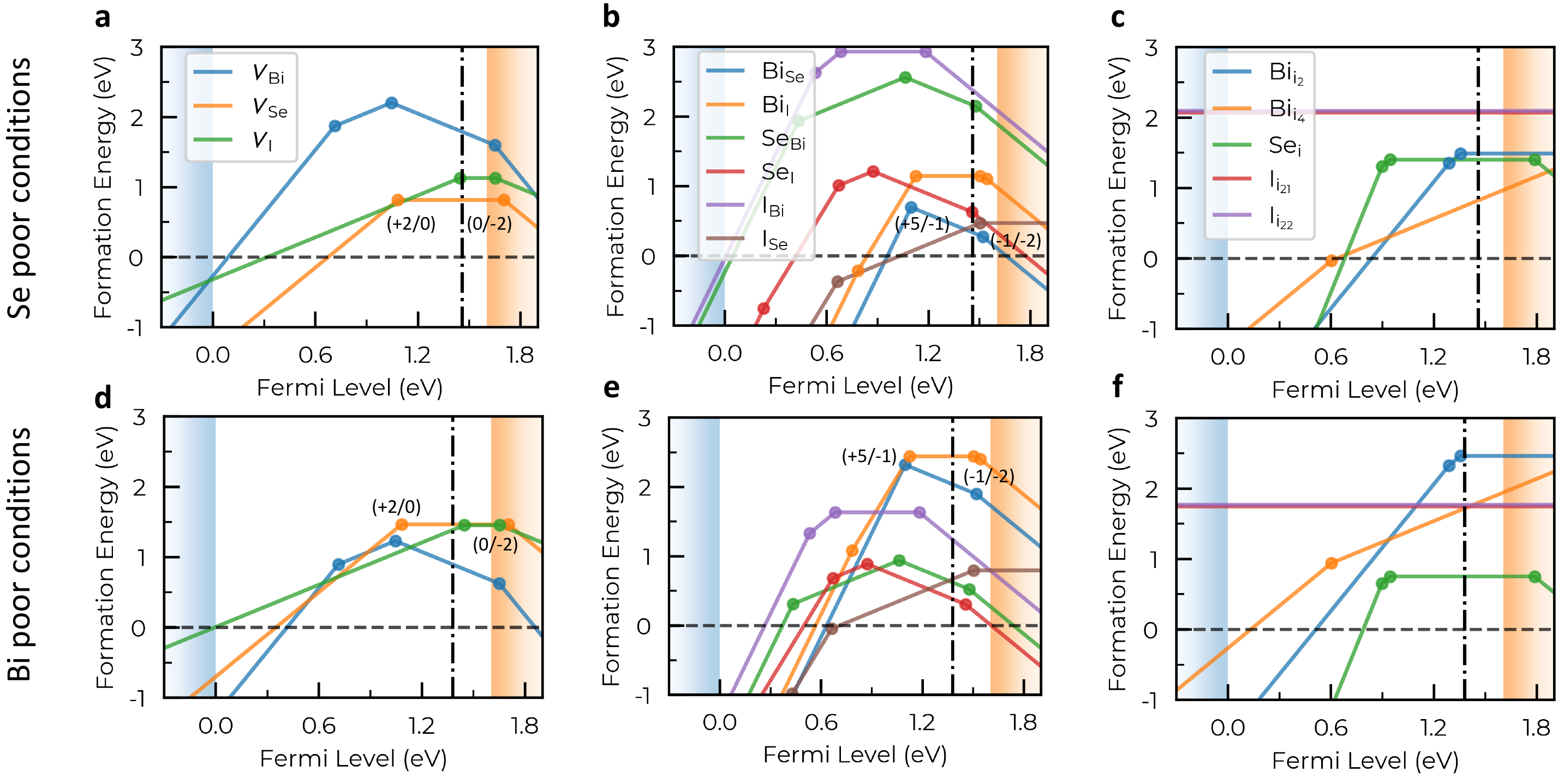}
        \caption{\textbf{Formation energies of point defects in BiSeI.} \textbf{a-c.}~Se-poor growth conditions. \textbf{d-f.}~Bi-poor growth conditions. The self-consistent Fermi level, $E^{\rm sc}_{F}$, is represented with vertical line-dot lines, lying at $1.46$ and $1.38$~eV above the VBM (blue shaded region) for Se-poor and Bi-poor growth conditions, respectively (the CBM is represented by the orange shaded region). Numerical subscripts in panels c and f indicate inequivalent lattice interstitial positions (Supplementary Fig.~1).}
        \label{Figure: BiSeI diagram}
\end{figure*}

\textbf{Defect formation energies.}~Our defect formation energy ($E_{\rm f}$) results, expressed as a function of the Fermi level (VBM $\le E_F \le$ CBM), are shown in Fig.~\ref{Figure: BiSeI diagram}. These defect formation energies depend on the synthesis conditions (i.e., atomic chemical potentials), defect charge state, and Fermi energy level (Methods). A charge-state transition occurs when the energy curves of two different charge states intersect, signaling the potential exchange of charge carriers between the defect and the host material. Defects with low formation energies can significantly impact electron-hole recombination processes, potentially reducing the material’s PCE. This effect is particularly detrimental when transition energy levels are deep within the bandgap, as opposed to those near the band edges \cite{Freysoldt2014}.

The chemical potentials of BiSeI are bounded by two limiting cases: Se-poor conditions \((\mu_{\rm Bi}, \mu_{\rm Se}, \mu_{\rm I}) = (0, -0.65, -0.75)\)~eV and Bi-poor conditions \((\mu_{\rm Bi}, \mu_{\rm Se}, \mu_{\rm I}) = (-0.97, 0, -0.42)\)~eV (Fig.~\ref{Figure: BiSeI structure}c), where the competing secondary phases are Bi-Bi$_{2}$Se\textsubscript{3} and Se-BiI\textsubscript{3}, respectively. Figure~\ref{Figure: BiSeI diagram} illustrates the \(E_{\rm f}\) results obtained for these two extreme synthesis conditions. Intermediate synthesis conditions can be explored using the open access data provided in \cite{database} and Supplementary Tables~2--7. Regarding experimental synthesis, Se-poor conditions are commonly encountered in physical synthesis routes due to the high volatility of selenium atoms at elevated temperatures \cite{Cano2023,Li2024}. 

For many defects, the calculated formation energies are very low under Se-poor synthesis conditions (Figs.~\ref{Figure: BiSeI diagram}a--c), yielding high defect concentrations. Bi-poor conditions, on the other hand, favor moderate to high defect formation energies (Fig.~\ref{Figure: BiSeI diagram}d--f). Notable examples include: (i)~V$_{\rm Se}$, which exhibits a formation energy of \(0.82\)~eV at the self-consistent Fermi level, $E^{\rm sc}_{F}$, under Se-poor conditions compared to \(1.47\)~eV under Bi-poor (Figs.~\ref{Figure: BiSeI diagram}a,d), and (ii)~Bi$_{\rm Se}$, with \(E_{f} = 0.34\)~eV at $E^{\rm sc}_{F}$ under Se-poor conditions and \(2.04\)~eV under Bi-poor (Figs.~\ref{Figure: BiSeI diagram}b,e).

In terms of photovoltaic (PV) performance, the most detrimental defects are those with low formation energies at the self-consistent Fermi level \cite{Ganose2018}. $E^{\rm sc}_{F}$ is the equilibrium Fermi level that ensures a zero-charge balance across the defect and carrier populations in the system \cite{Squires2023}, and it can vary with temperature (e.g., higher temperatures induce larger defect/carrier populations) and chemical potentials (i.e., growth conditions). Therefore, exploration of the self-consistent Fermi level under different temperatures is essential to accurately assess the defects energy and its impact on PV performance.

Under Bi-poor synthesis conditions and at room temperature, $E^{\rm sc}_{F}$ is positioned $1.38$~eV above the VBM (considering defect concentrations generated at a realistic annealing temperature of $550$~K). Similarly, under Se-poor growth conditions, $E^{\rm sc}_{F}$ lies $1.46$~eV above the VBM. In both cases, $E^{\rm sc}_{F}$ exhibits a weak dependence on annealing (Supplementary Fig.~2), which indicates a marked $n$-type character for BiSeI. This behavior is consistent with previous experimental works showing that $p$-type doping is very challenging in MChX \cite{Cano2023}.  

In Se-poor environments, several antisite defects present very low formation energies at $E^{\rm sc}_{F}$, the most critical cases being  I\textsubscript{Se} (0.43~eV) and Bi\textsubscript{Se} (0.34~eV) (Fig.~\ref{Figure: BiSeI diagram}b). Bi-poor synthesis conditions generally result in higher formation energies; however, certain antisite defects still exhibit very low $E_{\rm f}$ (Fig.~\ref{Figure: BiSeI diagram}e): I\textsubscript{Se} (0.68~eV), Se\textsubscript{I} (0.38~eV), and Se\textsubscript{Bi} (0.62~eV). Among these, Bi\textsubscript{Se} stands out as the most detrimental antisite, acting as a potential \textit{killer} defect under typical experimental synthesis conditions. This defect undergoes a pronounced geometric reconstruction during its $(+5/-1)$ charge-state transition, transforming from a true substitutional defect in the $-1$ state to a defect complex, Bi\textsubscript{i} + V\textsubscript{Se}, in the $+5$ state (Supplementary Fig.~3).

Formation energies of vacancies (Figs.~\ref{Figure: BiSeI diagram}a,d) and interstitials (Figs.~\ref{Figure: BiSeI diagram}c,f) commonly are less favorable than those of antisites. The quasi-one-dimensional structure of BiSeI plays a significant role in this $E_{\rm f}$ trend. Point defects are predominantly confined within the atomic columns where they originate, minimizing interference between neighboring columns (Supplementary Fig.~4). Vacancies, which require significant structural adjustments within the affected column, tend to exhibit moderate formation energies. Interstitials, positioned between adjacent columns, are even more challenging to create due to the extensive lattice disruptions that they induce. In contrast, antisite defects --where one atom is replaced by another within the same column-- necessitate minimal lattice reorganization and consequently are easier to form. These trends in defect formation are consistent with observations in columnar pnictogen chalcogenides (e.g., Sb\textsubscript{2}S\textsubscript{3} and Sb\textsubscript{2}Se\textsubscript{3} \cite{Savory2019, Wang2023}).

Among all vacancies, V$_{\rm Se}$ has the lowest formation energy at $E^{\rm sc}_{F}$, with a value of $0.82$~eV under Se-poor growth conditions. This outcome suggests that V$_{\rm Se}$ could act as a nonradiative charge-recombination center, potentially contributing to PCE losses.
\\

\begin{figure*}[t]
    \centering
        \includegraphics[width=\textwidth]{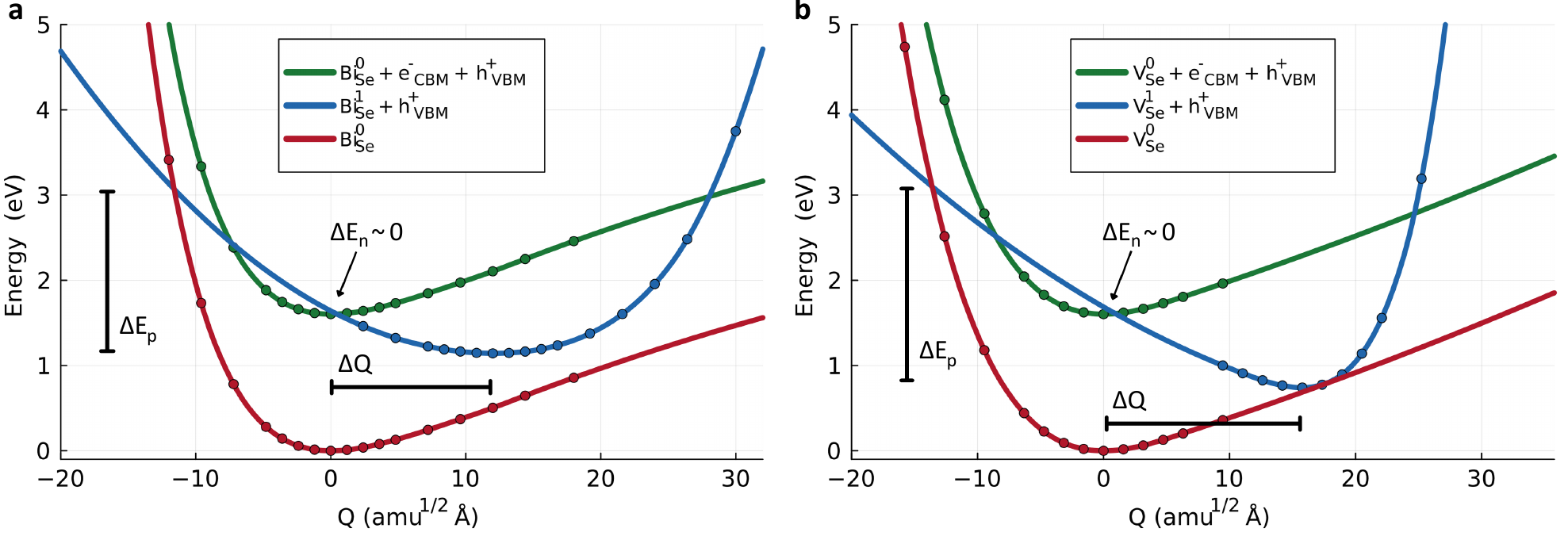}
        \caption{\textbf{Configuration coordinate diagrams for BiSeI.}
        Configuration coordinate diagrams for \textbf{a.}~Bi$_{\rm Se}$ $(0/+1)$ and \textbf{b.}~V$_{\rm Se}$ $(0/+1)$. The dots represent potential energies explicitly computed from first principles, and the solid lines are their corresponding quadratic spline interpolation and extrapolation. $\Delta Q$ represents the generalized distance between charge states and $\Delta E_{x}$ the electron~(n)/hole~(p) energy capture barriers.}
        \label{Figure: BiSeI carrier-capture}
\end{figure*}

\textbf{Polaron formation.}~Polarons --localized charges accompanied by significant lattice distortions due to strong electron-phonon coupling-- can significantly impact carrier recombination via self-trapping mechanisms, potentially limiting PCE \cite{Kavanagh2021-2}. They are generally classified as small or large, depending on their spatial extent and interaction with the lattice. Small polarons are particularly detrimental, as they strongly enhance electron-hole recombination \cite{Quirk2023}. To assess this possibility, we analyzed carrier self-trapping phenomena induced by defect-bound polarons (Supplementary Discussion). Such processes may be pronounced in BiSeI due to the strong lattice distortions associated with its most critical defects, particularly Bi\textsubscript{Se}, facilitated by the anharmonic ionic-covalent bonding typical of lone-pair chalcogenides and chalcohalides. However, our calculations reveal weak electron–phonon coupling in BiSeI (Supplementary Discussion), indicating that charge carriers are unlikely to localize via phonon interactions. As a result, polarons in BiSeI are expected to be large, and can reasonably be ruled out as a primary factor limiting the PCE of MChX.
\\

\begin{figure*}[t]
    \centering
        \includegraphics[width=0.9\textwidth]{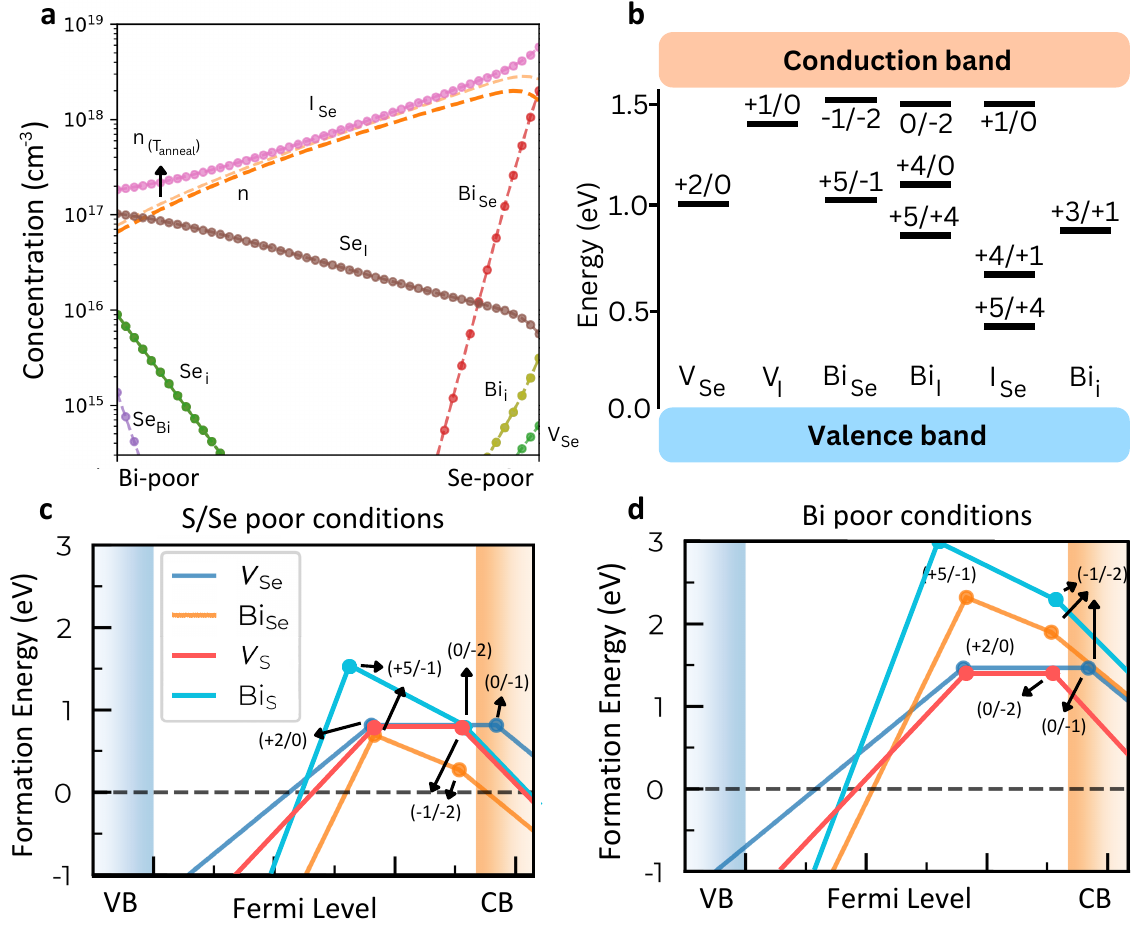}
        \caption{\textbf{Point defect chemistry of BiSeI and BiSBr.} \textbf{a.}~Defect concentrations of BiSeI considering an annealing temperature of $550$~K. \textbf{b.}~Sketch of the most prominent (i.e., lowest energy) point defects determined for BiSeI. Defect formation energies for BiSeI and BiSBr evaluated under \textbf{c.}~S/Se-poor and \textbf{d.}~Bi-poor growth conditions. The formation energy of Bi\textsubscript{S} is significantly higher than that of Bi\textsubscript{Se}.}
        \label{Figure: BiSBr antisite}
\end{figure*}

\textbf{Charge-carrier capture coefficients.}~Given their low formation energies at the self-consistent Fermi level, V\textsubscript{Se} and Bi\textsubscript{Se} are likely the most detrimental defects for PCE among all vacancies and antisites (interstitial defects are not considered further due to their higher $E_{\rm f}$ values). However, the PV performance of MChX is not solely dictated by defect formation energies. Electron-phonon coupling, particularly its role in electron/hole capture processes that drive nonradiative charge-carrier recombination, may also be critical \cite{Kim2020-2}. To address this aspect, we analyzed the impact of electron-phonon coupling on charge-carrier capture events associated with the V\textsubscript{Se} $(+2/0)$ and Bi\textsubscript{Se} $(+5/-1)$ charge-state transitions, following the computational methodology introduced in previous studies \cite{Alkauskas2014,Kim2019,Kim2020-2}.  

In a nutshell, the potential energy surface (PES) of the defect transition is first mapped along the structural path connecting the equilibrium geometries of the two defect charge states involved in the capture process (i.e., the $Q$ coordinate in Figs.~\ref{Figure: BiSeI carrier-capture}a,b). To accomplish this, multi-electron transitions are decomposed into sequential single-electron transition processes, with capture coefficients computed separately for each step [e.g., (+2/+1) and (+1/0) for V\textsubscript{Se} (+2/0)] \cite{Alkauskas2014}. Next, from the PES mapping, nuclear wavefunction overlaps are obtained by solving a one-dimensional Schrödinger equation \cite{Alkauskas2014,Kim2019}. Electron-phonon coupling is then evaluated using static perturbation theory. Finally, combining these results with phonon overlaps and scaling factors that account for charge interaction effects, the carrier capture coefficients are determined \cite{Kim2020-2} (Methods).  

For both V\textsubscript{Se} and Bi\textsubscript{Se} charge-state transitions, the most detrimental single-electron capture process for PV performance occurs at the (0/+1) level, as it exhibits the highest electron/hole capture coefficients ($C_{n/p}$, Supplementary Table~8). Therefore, we focus on this particular process here. Since the estimated $C_{n/p}$ coefficients show minimal temperature dependence (Supplementary Fig.~5), we neglect this effect in the following discussion.  

The energy barriers for (0/+1) electron and hole capture processes, $\Delta E_{n/p}$, are critical parameters in determining $C_{n/p}$ (Methods). In BiSeI, the electron capture barriers ($\Delta E_{n}$, Figs.~\ref{Figure: BiSeI carrier-capture}a,b) are found to be nearly negligible for both V\textsubscript{Se} and Bi\textsubscript{Se}. In contrast, the hole capture barriers ($\Delta E_{p}$, Figs.~\ref{Figure: BiSeI carrier-capture}a,b) are significant, amounting to $2.36$~eV and $1.93$~eV for V\textsubscript{Se} and Bi\textsubscript{Se}, respectively. These $\Delta E_{n/p}$ results suggest high $C_{n}$ and low $C_{p}$ for MChX. 

Although this expectation holds for Bi\textsubscript{Se} ($C_{n} = 6.20 \cdot 10^{-11}$~cm\textsuperscript{3}/s and $C_{p} < 10^{-20}$~cm\textsuperscript{3}/s), it does not hold for V\textsubscript{Se}. Specifically, the room-temperature hole capture coefficient estimated for V\textsubscript{Se} is notably high, reaching $9.85 \cdot 10^{-8}$~cm\textsuperscript{3}/s, while its electron capture coefficient is $1.91 \cdot 10^{-10}$~cm\textsuperscript{3}/s (Supplementary Fig.~5, Supplementary Table~8). This unexpected result stems from factors beyond $\Delta E_{n/p}$, such as electron-phonon coupling and phonon wave function overlaps, which also play a critical role in carrier capture processes (Methods).  

The capture coefficient values estimated for BiSeI are relatively small compared to those found in similar materials such as Sb\textsubscript{2}Se\textsubscript{3} ($C_{n} = 5.63 \cdot 10^{-6}$~cm\textsuperscript{3}/s and $C_{p} = 1.22 \cdot 10^{-8}$~cm\textsuperscript{3}/s) \cite{Wang2024}. At first glance, this comparison suggests that defect chemistry may have a notable, though less pronounced, impact on the PV efficiency of BiSeI compared to Sb\textsubscript{2}Se\textsubscript{3}. However, as we will discuss in the next section, this expectation does not hold.

\section*{Discussion} 
The detailed balance model \cite{Shockley1961}, which neglects nonradiative charge-carrier recombination losses, predicts a maximum PCE of 30.47\% for BiSeI at room temperature, along with an open-circuit voltage (V\textsubscript{oc}) of $1.33$~V and a fill factor (FF) of 90.54\% \cite{Lopez2024}. This ideal Shockley–Queisser limit assumes that each absorbed photon generates an electron–hole pair and that all charge-carrier recombination is purely radiative. When thickness-dependent absorptivity is considered, the radiative efficiency limit for a $700$~nm-thick absorber layer slightly decreases to 30.39\%, accompanied by a reduced $V_{\text{oc}}$ of $1.33$~V and a nearly unchanged FF of 90.54\%. These results underscore the potential of BiSeI for photovoltaic applications, as its power conversion efficiency is only minimally affected by finite-layer thickness.

However, as noted earlier, PV cells based on MChX absorbers have yet to exceed a PCE of 10\% \cite{Cano2023, Nie2021, Tiwari2019}. While factors such as material morphology and device architecture may contribute to this limitation, our findings suggest that trap-mediated nonradiative charge recombination could also play a significant role in reducing PV efficiency.  

The maximum defect-limited PCE of BiSeI, $\eta$, can be estimated using the calculated $C_{n/p}$, defect concentrations, and related parameters (Methods) \cite{Dahan2013}. Under Se-poor synthesis conditions and assuming an annealing temperature of 550~K, we estimate the power conversion efficiency (PCE), $\eta$, to be 24.2\% --a notable reduction of over 6\% compared to the ideal detailed balance limit. Separate $\eta$ calculations for Bi\textsubscript{Se} and V\textsubscript{Se} defects reveal that this efficiency loss originates entirely from the selenium vacancy. In fact, the maximum PCE estimated when considering only the antisite defect remains nearly identical to the ideal limit, owing to its extremely low carrier capture coefficients. Therefore, substitutional traps such as Bi\textsubscript{Se}, and similarly I\textsubscript{Se} (Figs.~\ref{Figure: BiSeI diagram}b,e), can be regarded as electronically benign and ruled out as significant nonradiative recombination centers in BiSeI.

The nonradiative efficiency loss in BiSeI is comparable to that observed in other light absorber materials, such as Cu\textsubscript{2}ZnSnS\textsubscript{4} \cite{Kim2021}, Cu\textsubscript{2}ZnSnSe\textsubscript{4} \cite{Kim2020-2}, and CdTe \cite{Kavanagh2021}. Additionally, the PCE loss in BiSeI is accompanied by a notable reduction in open-circuit voltage (V\textsubscript{oc} = 1.08~V) and fill factor (FF = 88.81\%), indicating a deterioration in electronic quality under illumination and reduced charge transport efficiency.  

The estimated $\eta$ for BiSeI is slightly lower than that of Sb\textsubscript{2}Se\textsubscript{3}, which has a predicted defect-limited efficiency of 26\% \cite{Wang2024}. At first glance, this result may seem counterintuitive, as the calculated carrier capture coefficients for Sb\textsubscript{2}Se\textsubscript{3} are significantly higher than those for BiSeI. However, the discrepancy is explained by the lower defect formation energies in BiSeI (e.g., $1.2$~eV for V\textsubscript{Se} in Sb\textsubscript{2}Se\textsubscript{3} \cite{Wang2024} versus $0.8$~eV for the same defect in BiSeI), which lead to a substantially higher concentration of defects and free carriers (Fig.~\ref{Figure: BiSBr antisite}a). This finding highlights that PCE limitations in MChX semiconductors arise primarily from the high abundance of defects, rather than their individual recombination strengths. As a result, effective defect passivation strategies could play a critical role in improving the photovoltaic performance of these materials.
 
Based on our first-principles computational results, several strategies can be proposed to mitigate the formation of PV-detrimental defects in MChX. One promising approach is optimizing synthesis conditions. In particular, our study demonstrates that adopting Bi-poor growth conditions --despite the associated practical challenges \cite{Cano2023}-- significantly increases the formation energy of most defects (Fig.~\ref{Figure: BiSeI diagram}), thereby reducing their prevalence. Notably, under Bi-poor synthesis conditions, our calculated $\eta$ reaches 30.39\%, remaining very close to the ideal detailed balance limit, with an open-circuit voltage of 1.33~V and a fill factor of 90.54\%.  

Another approach to mitigating PCE losses is ion substitution, which provides a practical and controlled method for engineering defect chemistry (Fig.~\ref{Figure: BiSBr antisite}b). The similar ionic radii of Bi ($r_{\rm Bi} = 207$~pm), Se ($r_{\rm Se} = 190$~pm), and I ($r_{\rm I} = 198$~pm) facilitate their interchangeability and diffusion at elevated temperatures, promoting the formation of antisite and vacancy defects. To address this issue, we propose chemically guided ion substitutions involving atomic species with larger ionic radius differences, while preserving the desirable optoelectronic properties. Based on this principle, BiSBr emerges as a promising MChX candidate for reduced defect concentrations, benefiting from the substantial size mismatch among Bi ($r_{\rm Bi} = 207$~pm), S ($r_{\rm S} = 180$~pm), and Br ($r_{\rm Br} = 183$~pm).  

Supplementary defect calculations confirm that the some defects in BiSeI can be significantly passivated through compositional substitution in BiSBr (Figs.~\ref{Figure: BiSBr antisite}c,d). Specifically, under S-poor synthesis conditions, the charge-state transition Bi\textsubscript{S} ($-1/-2$) exhibits a formation energy of 0.80~eV, compared to only 0.28~eV for Bi\textsubscript{Se} in BiSeI (Fig.~\ref{Figure: BiSBr antisite}c). Similarly, under Bi-poor conditions, the formation energy of Bi\textsubscript{S} ($-1/-2$) increases to 2.30~eV, while that of Bi\textsubscript{Se} remains lower at 1.90~eV (Fig.~\ref{Figure: BiSBr antisite}d). In contrast, the formation energy of sulfur and selenium vacancy charge-state transitions remain nearly unchanged across the two compounds, with Bi-poor conditions being generally the most favorable for defect suppression (Figs.~\ref{Figure: BiSBr antisite}c,d). Overall, the Se$\to$S and I$\to$Br substitutions preserve the charge-state transition levels of vacancy defects, while increasing the formation energies of antisite defects by more than 200\%. This substantial enhancement contributes to the improved defect tolerance of BiSBr compared to BiSeI.

Finally, it is hypothesized that additional ion substitutions could further enhance defect mitigation in MChX materials \cite{Nicolson2023} by reducing antisite and vacancy concentrations. For instance, BiOI, which exhibits a significant size difference between O ($r_{\rm O} = 152$~pm) and I ($r_{\rm I} = 198$~pm), has a Bi–O ionic radius difference of 55~pm, that is, more than 220\% larger than the Bi–Se difference in BiSeI. Similarly, the I–O ionic radius difference of 46~pm represents a 475\% increase compared to the I–Se difference in BiSeI. Notably, BiOI has recently emerged as a promising candidate for both photocatalysis \cite{Dai2016} and photovoltaic absorption \cite{Sfaelou2015, Feeney2023}, suggesting that it may exhibit great defect tolerance in addition to excellent optoelectronic properties.

\section*{Conclusions}
This study identifies the critical role of point defects in limiting the PV performance of MChX light absorbers, with BiSeI as model system. Using advanced first-principles and configuration sampling methods, we show that V\textsubscript{Se}, followed by Bi\textsubscript{Se}, is the most detrimental defect, significantly facilitating nonradiative charge-carrier recombination. These defects reduce the maximum PCE of BiSeI down to approximately $24$\%, which is about 6\% smaller than the corresponding ideal detailed balance limit. This performance loss is accompanied by reductions in open-circuit voltage (V\textsubscript{oc} = $1.08$~V) and fill factor (FF = 88.81\%). Nevertheless, we propose potential defect mitigation strategies based on synthesis conditions and chemical substitutions. Specifically, Bi-poor growth conditions significantly increase the formation energies of V\textsubscript{Se} and Bi\textsubscript{Se}, reducing their concentrations. Additionally, the ion substitutions S→Se and Br→I offer promising improvements in defect tolerance. 

The findings presented in this study are significant for the field of photovoltaics, as they provide a pathway to enhance the efficiency of MChX light absorbers, a promising class of nontoxic and thermodynamically stable materials. By addressing defect chemistry and proposing effective passivation strategies, this study not only bridges the gap between theoretical predictions and experimental performance but also establishes a framework for designing more efficient solar absorbers with improved defect tolerance. These insights could pave the way for next-generation photovoltaic technologies with higher efficiencies and broader applicability.
\\

\section*{Methods}

\textbf{First-principles calculations outline.}~\textit{Ab initio} calculations based on density functional theory (DFT) were  performed to analyse point defects in MChX. These calculations were conducted with the VASP code \cite{Kresse1996} using the generalized gradient approximation to the exchange-correlation energy for solids due to Perdew \textit{et al.} (PBEsol) \cite{Perdew1996}. Since MChX are van der Waals materials, long-range dispersion interactions were taken into account through the D3 scheme \cite{Grimme2010}. The projector augmented-wave method was used to represent the ionic cores \cite{Blochl1994} and for each element the maximum possible number of valence electronic states was considered. Wave functions were represented in a plane-wave basis typically truncated at $600$~eV. By using these parameters and a dense \textbf{k}-point grid for reciprocal space Brillouin zone integration of $11 \times 5 \times 4$ (centered at $\Gamma$), the resulting energies were converged to within $1$~meV per formula unit. In the geometry relaxations, a tolerance of $0.5$~meV$\cdot$\AA$^{-1}$ was imposed in the atomic forces. Defect calculations were performed in $3 \times 2 \times 1$ ($12.8 \times 18.1 \times 11.3$~\AA) supercells, using a $2 \times 1 \times 2$ $\Gamma$-centered \textbf{k}-point grid. For the estimation of optoelectronic properties (e.g., band gaps and optical absorption coefficients), spin-orbit coupling corrections were taken into account along with range-separated hybrid functionals containing an exact Hartree-Fock exchange fraction of $25$\% (i.e., HSEsol+SOC \cite{Heyd2003,Krukau2006}).

The static dielectric tensor ($\varepsilon_{stat}$) included both electronic ($\varepsilon_{\infty}$) and ionic ($\varepsilon_{0}$) contributions. For the computation of the ionic (lattice response) contribution, the dynamical matrix was constructed from a finite-differences approach, employing the PBEsol functional. For the electronic (optical response) contribution, the HSEsol+SOC functional was used, with a $8 \times 4 \times 3$ $\Gamma$-centered \textbf{k}-point mesh (Supplementary Table~9).
\\

\textbf{Exploration of the potential energy surface.}~Conventional approaches to generating defect configurations \cite{Freysoldt2014} from pristine cells fail to find many lowering-energy conformations, which might have a crucial effect in the conclusions. Therefore, once a defect is generated (e.g., extracting a bismuth atom), we look for distortions of the initial lattice configuration to locally explore the potential energy surface. These distortions were generated with ShakeNBreak approach \cite{Mosquera-Lois2022,Mosquera-Lois2023}.

Initially, all the trial defect configurations were relaxed using $\Gamma$-point reciprocal space sampling and the HSEsol+D3 functional. Only the minimum-energy configurations were kept. Next, the ionic relaxations were repeated considering a larger \textbf{k}-point grid of $2 \times 1 \times 2$ ($\Gamma$-centered). After that, the relaxations were repeated considering SOC corrections. Finally, single-point energy calculations were performed using the previously-converged electronic wavefunctions and equilibrium structures. Non-spherical contributions to the gradient of the density within the PAW spheres were taken into account to improve numerical accuracy.
\\

\textbf{Point-defect formation energies.}~The formation energy of a point defect with charge $q$, $D^{q}$, can be expressed as \cite{Freysoldt2009}:
\begin{equation}
    \begin{gathered}
        E_{\rm f}(D^q) = E_T(D^q) - E_T({\rm pristine}) + \\
         + q E_F - \sum_i n_i \mu_i + E_{corr} (D^q)~,
    \end{gathered}
    \label{Eq. defect formation energy}
\end{equation}
where $E_T(D^q)$ and $E_T({\rm pristine)}$ are the static energies of defected and pristine supercells (energy per formula unit), respectively, $\mu_i$ corresponds to chemical potential of species $i$ (this is, the energy required to extract one single atom), $n_i$ the number of  extracted atoms (positive or negative if extracted or added to the pristine cell, respectively), $E_F$ the Fermi energy (energy needed to extract an electron), and $E_{corr}$ the finite-size corrections based on spurious interactions between charged defects due to the periodic boundary conditions.

Here we considered two different contributions to the finite-size energy correction: point-charge (due to the spurious electrostatic interactions of a defect with its images) and band-alignment corrections (charged defects spuriously change the electrostatic potential of the system). Both corrections are computed together from an extension of the Freysoldt-Neugebauer-Van de Walle \cite{Freysoldt2009} correction scheme to anisotropic materials \cite{Kumagai2014}, as implemented in the \verb!doped! defect simulation package \cite{Kavanagh2024} (Supplementary Tables~2--7).
\\

\textbf{Fröhlich coupling constant.}~The Fröhlich (or polaron) coupling constant for electron and holes reads:
\begin{equation}
    \alpha^F_{(e, h)} = \frac{e^2}{4 \pi \epsilon_0 \hbar} \left( \frac{1}{\varepsilon_{\infty}} - \frac{1}{\varepsilon_{stat}} \right) \sqrt{ \frac{m_{(e, h)}}{2 \hbar \omega_{LO}}}~,
\end{equation}
where $\epsilon_{0}$ is the vacuum permittivity, $\hbar$ is the reduced Planck’s constant, $m_{(e, h)}$ is the effective mass of electron or hole, and $\omega_{LO}$ is the effective longitudinal optical phonon frequency taken as the average over all $\Gamma$-point modes weighted by the dipole moment they produce \cite{Ganose2021} (additional technical details can be found in the Supplementary Discussion).
\\

\textbf{Defect-limited efficiency.}~The nonradiative recombination activity has been estimated from the electron and hole capture coefficients for each charge state of the defect, using the \verb!CarrierCapture.jl! package \cite{Alkauskas2014,Kim2020}.

Within the employed formalism, the maximum defect-limited photovoltaic efficiency \cite{Dahan2013} under incident radiation spectrum $\Phi$ is:
\begin{equation}
    \eta = \max_V \left( \frac{J V}{q \int_0^{\infty} E \Phi(E) dE} \right)~,
\end{equation}
where $q$ is the electron charge and $J$ the maximum defect-limited current density:
\begin{equation}
    J (W, V) = J_{SC} (W) + J_0^{rad} (W, V) + J_0^{nonrad} (W, V)~,
\end{equation}
$J_{SC}$, $J_0^{rad}$ are the short-circuit and saturation currents, respectively. While these two terms lead the annihilation of charges due to radiative recombination, $J_0^{nonrad}$ takes into account the non-radiative recombination:
\begin{equation}
    J_{SC} (W) = q \int_0^{\infty} a (E, W) \Phi (E) dE
\end{equation}
\begin{equation}
    \begin{gathered}
        J_0^{rad} (W, V) = q \frac{2 \pi}{c^2 h^3} \left( 1 - e^{\frac{q V}{k_B T}} \right) \times \\
        \times \int_0^{\infty} a (E, W) \left( e^{\frac{E}{k_B T}} - 1 \right)^{-1} E^2 dE
    \end{gathered}
\end{equation}

\begin{equation}
    J_0^{nonrad} (W, V) = - q W R^{SRH} (V)~,
\end{equation}
where $a$ is the absorptivity of the system (detailed balance limit assuming that each photon generates an electron-hole pair) and:
\begin{equation}
    R_{SRH} \approx \sum \Delta n/p ~N_T C_{n/p}~,
    \label{Eq. recombination rate}
\end{equation}
where $\Delta n$, $\Delta p$, and $N_{T}$ denote excess concentrations of electrons, holes, and concentration of defects, respectively, with the summation over all independent recombination centers in the system. The carrier capture coefficient \cite{Kim2020-2,Alkauskas2014} ($C_{n/p}$) can be expressed using the electron-phonon coupling ($W{ct}$) and the overlap of phonon wave functions ($\langle \zeta_{cm} | \Delta Q | \zeta_{tn} \rangle$), which is given by:
\begin{equation}
    \begin{gathered}
        C_{n/p} = \Omega \frac{2 \pi}{\hbar} |W_{ct}|^2 \sum_{m,n} w_m \langle \zeta_{cm} | \Delta Q | \zeta_{tn} \rangle^2 \times \\
        \times \delta(\Delta E_{n/p} + \epsilon_{cm} - \epsilon_{tn})~,
    \end{gathered}
    \label{Eq. carrier capture}
\end{equation}
with $\Omega$ the volume of the supercell, $g$ the degeneracy of the defect, $\zeta$ the phonon wave function, $\Delta Q$ an effective configuration coordinate for the phonon wave functions, and the subscripts $c$ and $t$ the free carrier and trap states, respectively. In this formalism, the temperature-dependence is determined by the thermal occupation number $w_m$ of the initial vibrational state. The TLC package \cite{Kim2020-2} was used to estimate the effect of carrier-capture kinetics on photovoltaic efficiency.
\\

\section*{Data availability}
The data that support the findings of this study have been made publicly available \cite{database}, comprising the single-point energy and local potential calculations for all relaxed defects.
\\

\section*{Acknowledgments}
C.L. acknowledges support from the Spanish Ministry of Science, Innovation and Universities under a FPU grant. S.R.K. acknowledges the Harvard University Center for the Environment (HUCE) for funding a fellowship. C.C. acknowledges support by MICIN/AEI/10.13039/501100011033 and ERDF/EU under the grants TED2021-130265B-C22,
TED2021-130265B-C21, PID2023-146623NB-I00, PID2023-147469NB-C21 and RYC2018-024947-I and by the Generalitat de Catalunya under the grants 2021SGR-00343, 2021SGR-01519 and 2021SGR-01411. Computational support was provided by the Red Española de Supercomputación under the grants FI-2024-1-0005, FI-2024-2-0003 and FI-2024-3-0004. This work is part of the Maria de Maeztu Units of Excellence Programme CEX2023-001300-M funded by MCIN/AEI (10.13039/501100011033). P.B. acknowledges support from the predoctoral program AGAUR-FI ajuts (2024 FI-1 00070) Joan Oró. E.S. acknowledge the European Union H2020 Framework Program SENSATE project: Low dimensional semiconductors for optically tuneable solar harvesters (grant agreement Number 866018), Renew-PV European COST action (CA21148), the Spanish Ministry of Science and Innovation ACT-FAST (PCI2023-145971-2), and the ICREA Academia Program.
\\

\section*{Author contributions}
C.L., E.S. and C.C. conceived the study and planned the research, which was discussed in-depth with the rest of the co-authors. C.L., S.K. and P.B. performed the first-principles calculations and analyzed the results. The manuscript was written by C.L. and C.C., with substantial input from the rest of the co-authors.
\\

\section*{Additional information}
Supplementary information is available in the online version of the paper.
\\

\section*{Competing financial interests}
The authors declare no competing financial interests.
\\

%

\end{document}